\documentclass{jetpl}
\twocolumn
\title{Fermion zero modes at the boundary of superfluid $^3$He-B}

\rtitle{Fermion zero modes at the boundary of superfluid $^3$He-B}

\sodtitle{Fermion zero modes at the boundary of superfluid $^3$He-B}

\author{G.E. Volovik
 $^{\#}$\/\thanks{
volovik@boojum.hut.fi} 
}

\rauthor{G.E.Volovik}

\sodauthor{Volovik}

\address{Low Temperature Laboratory, Helsinki University of
Technology, P.O.Box 5100, FIN-02015, HUT, Finland
\\
 Landau Institute for Theoretical Physics RAS, Kosygina 2,
119334 Moscow, Russia}

\dates{July 30, 2009}{*}

\abstract{
Superfluid $^3$He-B belongs
to the important special class of  time-reversal invariant topological superfluids.
It has Majorana fermions as edge states on the 
surface of bulk $^3$He-B. On the rough wall these  fermion zero modes  have finite density of states at $E=0$.  It is possible that Lancaster experiments with a wire vibrating in $^3$He-B have already probed Majorana fermions  living on the surface of the wire.
}

\begin{document}

\maketitle


\section{Introduction}

General classification schemes based on topology \cite{Schnyder2008,Schnyder2009a,Schnyder2009b,Kitaev2009,Volovik2003,Volovik2007,Horava2005} suggest existence of the topological insulators and fully gapped topological superfluids which have the gapless edge states on the boundary. Superfluid $^3$He-B belongs to the topologically nontrivial class.
Topology of the spectrum of fermion zero modes at the interface between two bulk  $^3$He-B states with different realization of the order parameter has been discussed  in Ref. \cite{SalomaaVolovik1988}. In particular it was found that within some domain walls  the fermion zero modes have finite density of states at zero energy (Figure 12 in Ref. \cite{SalomaaVolovik1988}). The walls also have spin current carried by zero modes. 

Here we extend these results to the case of the fermion zero modes living on the boundary of $^3$He-B. We show that the problem of fermion zero modes on the wall with specular and with diffusive reflection is mapped to the problem of the  fermion zero modes living on different types of the interface  considered  in Ref. \cite{SalomaaVolovik1988}. In particular, the density of states on a perfect boundary is $N(E) \propto E$, while the density of states on a rough wall is finite, $N(E=0) \neq 0$.

\section{Hamiltonian}

Lets us consider the edge states on the boundary of $^3$He-B. The order parameter in superfluid $^3$He is $3\times 3$ matrix $A_{\alpha i}$. The simplest form of the order parameter in bulk $^3$He-B
 is $A_{\alpha i}
=\Delta_0\delta_{\alpha i}$, where $\Delta_0$ is the gap in the spectrum of fermionic quasiparticles \cite{VollhardtWolfle}.
The other states of the  bulk $^3$He-B are obtained by spin and phase rotations. 
 Near the wall the order parameter is distorted from its bulk value, and distortion is different
for the components parallel and perpendicular to the wall:
\begin{equation}
A_{\alpha i}= \left(
\begin{matrix}
 \Delta_\parallel(z)&0&0\cr
0&\Delta_\parallel(z)&0\cr
0&0&\Delta_\perp(z)\cr
\end{matrix} 
\right)~.
\label{OrderParameter}
\end{equation}
Here the normal to the surface is chosen along the axis $z$;   $\Delta_\perp(z=\infty)=\Delta_\parallel(z=\infty)=\Delta_0$.

Fermions are described by the following Hamiltonian:
\begin{eqnarray}
H=\frac{k^2-k_F^2}{2m^*} \tau_3+
\nonumber
\\
   \tau_1 \left(\Delta_\parallel(z) \sigma_x  \frac{k_x}{k_F}+ \Delta_\parallel(z)\sigma_y  \frac{k_y}{k_F} +\Delta_\perp(z)\sigma_z \frac{k_z}{k_F}  \right)\,,
\label{eq:Hamiltonian}
\end{eqnarray}
where $\tau_i$ and $\sigma_i$ are Pauli matrices of Bogolyubov-Nambu spin  and nuclear spin correspondingly. 

\section{Specular reflection}

We first consider the wall with specular reflection of quasiparticles. After conventional (not Andreev) reflection $k_z$ changes sign.  In quasiclassical Hamiltonian this  corresponds to the change of $ \Delta_\perp(z)$ to $- \Delta_\perp(z)$ after reflection. As a result the problem  transforms to finding the spectrum of the fermion bound states at the domain wall separating
two bulk $^3$He-B states  
\begin{equation}
A_{\alpha i}(z=+\infty)= \left(
\begin{matrix}
 \Delta_0&0&0\cr
0&\Delta_0&0\cr
0&0&\Delta_0\cr
\end{matrix} 
\right)~
\label{OrderParameterInterface+}
\end{equation}
and 
\begin{equation}
A_{\alpha i}(z=-\infty)=
 \left(
\begin{matrix}
 \Delta_0&0&0\cr
0&\Delta_0&0\cr
0&0&-\Delta_0\cr
\end{matrix} 
\right)~,
\label{OrderParameterInterface-}
\end{equation}
This is analogous to the bound states at the domain wall on which the mass which enters Dirac equation 
changes sign, $M(z)=-M(-z)$ \cite{SalomaaVolovik1988}.

For small $k_x^2+k_y^2\ll k_F^2$ one may use the perturbation theory. Introducing $k_z=k_F -i\partial_z$ one obtains the Hamiltoinian
 \begin{eqnarray}
H=H^{(0)}+H^{(1)}~,
\label{eq:HamiltonianExpansion}
\\
H^{(0)}=-i v_F \tau_3\partial_z + \tau_1  \sigma_z\Delta_\perp(z) ~,
\label{eq:H0}
\\
H^{(1)}=     \tau_1\frac{ \Delta_\parallel(z)}{k_F} \left(\sigma_x  k_x +\sigma_y  k_y  \right)\,,
\label{eq:H1}
\end{eqnarray}
 where $v_F=k_F/m^*$;  the terms  quadratic in  $k_x$, $k_y$ and $\partial_z$ are neglected. 

The zero-order  Hamiltonian \eqref{eq:H0} is supersymmetric, since  $\Delta_\perp(z)$ changes sign across the interface, and thus it has  solutions with exactly zero energy. There are two solutions corresponding two orientations of spin:
 \begin{eqnarray}
\Psi_+(z) \propto  \left(
\begin{matrix}
 1\cr
i\cr
\end{matrix} 
\right)_\tau
 \left(
\begin{matrix}
 1\cr
0\cr
\end{matrix} 
\right)_\sigma
 \exp\left(-\frac{1}{v_F} \int_0^z dz \Delta_\perp(z)\right) ~,
 \label{eq:ZeroEnergy+}
 \\
 \Psi_-(z) \propto  \left(
\begin{matrix}
 1\cr
-i\cr
\end{matrix} 
\right)_\tau
 \left(
\begin{matrix}
 0\cr
1\cr
\end{matrix} 
\right)_\sigma
 \exp\left(-\frac{1}{v_F} \int_0^z dz \Delta_\perp(z)\right) \,.
\label{eq:ZeroEnergy_}
\end{eqnarray}

The second order secular equation produces the  $2\times 2$ Hamiltonian constructed from the  
matrix elements of the Hamiltonian $H^{(1)}$ in \eqref{eq:H1}:  
\begin{equation}
\left(
\begin{matrix}
 H^{(1)}_{++}&H^{(1)}_{+-}\cr
H^{(1)}_{-+}&H^{(1)}_{--}\cr
\end{matrix} 
\right) 
=c \left(
\begin{matrix}
0&k_y+ik_x\cr
k_y-ik_x&0\cr
\end{matrix} 
\right) 
 \,,
\label{eq:Effective}
\end{equation}
This corresponds to the effective Hamiltonian for the `relativistic' fermion zero modes
 \begin{equation}
H_{\rm zm} = c \hat{\bf z} \cdot({\mbox{\boldmath$\sigma$}} \times {\bf k})~,
\label{eq:ModesH}
\end{equation} 
with the speed of light 
 \begin{equation}
c=\frac{ \int_{-\infty}^\infty dz \frac{\Delta_\parallel(z)}{k_F} \exp\left(- \frac{2}{v_F} \int_0^z dz' \Delta_\perp(z')\right)}
{ \int_{-\infty}^\infty dz   \exp\left(- \frac{2}{v_F} \int_0^z dz' \Delta_\perp(z')\right)} 
 \,.
\label{eq:SpeedOfLight}
\end{equation}
The particular case of the wall with $\Delta_\parallel(z)=\Delta_\perp(z)=\Delta_0$
has been considered in Ref. \cite{ChungZhang2009}. It corresponds to the interface with $\Delta_\parallel(z)=\Delta_0$ and $ \Delta_\perp(z)=\Delta_0{\rm sign}(z)$ and gives $c=\Delta_0/k_F$.

Existence of the fermion zero modes at the domain walls in $^3$He-B is supported by the  topological invariant  determined in the combined $({\bf k},z)$ space, see Eq.(25) in Ref. 
 \cite{SalomaaVolovik1988}. It is also expressed as the difference between topological charges of bulk $^3$He-B states on two sides of the wall,  see Eq.(12) in Ref. \cite{SalomaaVolovik1988}. 
The relation between topological charges for  bulk system and for its edge states is discussed in Chapter 22 of Ref.\cite{Volovik2003}.   

Since the components of the Bogoliubov-Nambu fermions are connected by complex conjugation, fermion zero modes described by Hamiltonian \eqref{eq:ModesH} are  Majorana fermions. The topology which protects the coarse-grained relativistic Majorana $Z_2$ fermions in 2+1 and in 3+1 dimensions is discussed in Ref. \cite{Horava2005}.

\section{Diffusive reflection}

If there are impurities on the wall, the reflection from impurity changes not only the component $k_z$  of the quasiparticle momentum, but also one of the components of ${\bf k}_\parallel=(k_x,k_y)$. Let us consider single impurity or imperfection on the wall, and  the trajectory of quasiparticle which is reflected by the impurity. We consider the process in which the quasiparticle with momentum ${\bf k}=(k_x,k_y,k_z)$ acquires  momentum  
$\bar{\bf k}=(-k_x,k_y,-k_z)$ after reflection. 
This corresponds to the change $\Delta_x$ to $- \Delta_x$ and $\Delta_\perp$ to $-\Delta_\perp$ after reflection. Then the problem  is mapped to that of the spectrum of the fermion bound states at the domain wall in which 
 $\Delta_x(z)=-\Delta_x(-z)$,  $\Delta_y(z)=\Delta_y(-z)$ and $\Delta_\perp(z)=-\Delta_\perp(-z)$; i.e. the wall separates  the $^3$He-B state   \eqref{OrderParameterInterface+}
and the $^3$He-B state
\begin{equation}
A_{\alpha i}(z=-\infty)=
 \left(
\begin{matrix}
- \Delta_0&0&0\cr
0&\Delta_0&0\cr
0&0&-\Delta_0\cr
\end{matrix} 
\right)~,
\label{OrderParameterInterface-diff}
\end{equation}
In this case the matrix elements of  $H^{(1)}$ in \eqref{eq:Effective} do not contain $k_x$.
This is because $\Delta_x(z)$ is antisymmetric and $ \int_{-\infty}^\infty dz  \Delta_x(z)  \exp\left(- \frac{2}{v_F} \int_0^z dz' \Delta_\perp(z')\right)=0$. As a result the spectrum of Majorana zero mode becomes $E=c|k_y|$. This spectrum corresponds to the effective Hamiltonian $H_{\rm zm} =c\sigma_z k_y$ (see Fig.~12 of Ref. \cite{SalomaaVolovik1988}). 

This spectrum can be extended to general $k_x$. For simplicity, let as choose the model wall in which  $\Delta_x(z)= \Delta_\perp(z)$. In this case the spin rotation about axis $y$  by angle $\phi$, where $\sin\phi=k_x/k_F$, and expansion    $k_z=k_F\cos\phi -i\partial_z$ lead
to the following Hamiltonian:
\begin{eqnarray}
H=H^{(0)}+H^{(1)}~,
\label{eq:HamiltonianExpansionDiff}
\\
H^{(0)}=-i v_F \cos\phi~\tau_3\partial_z + \tau_1  \sigma_z\Delta_\perp(z) ~,
\label{eq:H0Diff}
\\
H^{(1)}=     \tau_1\frac{ \Delta_y(z)}{k_F}  \sigma_y  k_y  \,.
\label{eq:H1Diff}
\end{eqnarray}
This gives the spectrum $E= c |k_y|$ with the ``speed of light'' which depends on $k_x$:
 \begin{equation}
c(k_x)=\frac{ \int_{-\infty}^\infty dz \frac{\Delta_y(z)}{k_F} \exp\left(- \frac{2}{v_F\cos\phi} \int_0^z dz' \Delta_\perp(z')\right)}
{ \int_{-\infty}^\infty dz   \exp\left(- \frac{2}{v_F\cos\phi} \int_0^z dz' \Delta_\perp(z')\right)} 
 \,.
\label{eq:SpeedOfLight}
\end{equation}

The important property of the spectrum $E(k_x,k_y)=c(k_x)|k_y|$ is that it leads to the finite density of states $N(E=0)\neq 0$ (see caption to Fig.~12 in Ref. \cite{SalomaaVolovik1988}). Returning back to the  case of the boundary of $^3$He-B, one obtains that for the perfect boundary, where the spectrum
of bound states is $E=c|{\bf k}_\parallel|$,  the density of states $N(E)\propto E$. For  the boundary with impurities the non-zero density of states $N(E=0)\neq 0$ arises which is proportional to the density of impurities. The finite density of bound states near the diffusive wall  has been obtained in  Ref. \cite{Kopnin1991} by the method of quasiclassical Green's functions for a special model for boundary conditions (see also recent paper \cite{Nagai2008}  and references therein).

 \section{Lancaster experiment and Majorana fermions}
 
It is important that for diffusive boundary one can always find the zero energy bound state with $k_\parallel\approx k_F$.
This may explain the  experiments with wire vibrating in $^3$He-B
\cite{CCGMP,CGPS,PicketTurbulence}. 

Due to gapless spectrum of fermion zero modes, there are always the low energy bound states, which are occupied even at very low temperature $T\ll \Delta_0$. 
Let us consider a wire moving with velocity ${\bf v}$ in superfluid. In the reference frame of the wire the 
velocity of superflow is ${\bf v}_{\rm s}=-{\bf v}$ far from the wire and reaches the maximum value ${\bf v}_{\rm s}=-2{\bf v}$ near the surface of the cylindrical wire. Due to the Doppler shift $E({\bf k})=E_0({\bf k}) +
 {\bf k}\cdot {\bf v}_{\rm s}$, the energy of the occupied zero mode state with $E_0({\bf k}) \approx 0$ may reach the maximum possible value  $E_{\rm zero~mode}=2vk_F$. This is the zero mode with ${\bf k}_\parallel= k_F \hat{\bf v}$, which may exist at diffusive boundary. Outside the wire the minimal energy of quasiparticles in the wire frame is $E_{\rm bulk}= \Delta_0 - vk_F$. 
 
 The fermion is able to escape from the occupied bound state to the bulk liquid if $E_{\rm zero~mode}> E_{\rm bulk}$. This becomes possible if the velocity of the wire exceeds the critical velocity $v>v_c=\Delta_0/3k_F$.  The same critical velocity is obtained if instead of escape of thermal Majorana fermions
 from the surace, we consider the pair creation: one fermion of the created pair is the Majorana fermion on the surface of the wire and the other one is the quasiparticle in bulk liquid. Such creation is possible due to vibration of the wire and is described by the draw-well mechanism, which is equivalent to the Gershtein-Zel'dovich \cite{GershteinZel'dovich1970} mechanism of electron-positron pair production in strong fields (see Refs. \cite{CalVolAlternatingField,Lambert} and Chapter 26.2 in \cite{Volovik2003}).
 
In Lancaster experiments  with wire vibrating in $^3$He-B, emission of quasiparticles has been detected  \cite{CCGMP,CGPS,PicketTurbulence} when the velocity amplitude of the wire exceeds the critical  velocity $v_c \approx\Delta_0/3k_F$.  This suggests that this emission occurs due to existence of the Majorana zero modes  with zero energy on the surface of the wire.

 \section{Conclusion}

The fermion zero modes (Majorana fermions) living near the rough walls of container with superfluid 
$^3$He-B provide the finite density of states at $E=0$. This corresponds to the finite DOS of fermion zero modes living on particular interfaces -- domain walls between bulk $^3$He-B states. It is possible that Lancaster experiments with a wire vibrating in $^3$He-B have already probed Majorana fermions  living on the surface of the wire.

It is a pleasure to thank M.V. Feigelman, A. Kitaev, N.B. Kopnin, A.W.W. Ludwig and M.A. Silaev for valuable discussions. This work is supported in part by the Russian Foundation
for Basic Research (grant 06--02--16002--a) and the
Khalatnikov--Starobinsky leading scientific school (grant
4899.2008.2).


\end{document}